\documentclass[a4paper,11pt]{article}
\pdfoutput=1 

\usepackage{jheppub} 

\usepackage[T1]{fontenc} 

\usepackage{amsmath,amssymb}
\usepackage{epsfig}  
\usepackage{graphicx}   
\usepackage{slashed}       
\usepackage{tikz}
\usepackage{amsmath}
\usepackage{tikz-feynman}
\usepackage{url}
\usepackage{color}
\usepackage{multirow}
\usepackage{comment}
\usepackage{amssymb}
\usepackage{orcidlink}
\usepackage{caption}
\usepackage{subcaption}

\usepackage[version=3]{mhchem}
\DeclareUnicodeCharacter{2212}{-}

\newcommand{\masqtv}{\ensuremath{m_a^2 \big|_{\rm TV}}}
\newcommand{\matv}{\ensuremath{m_a \big|_{\rm TV}}}
\newcommand{\masqfv}{\ensuremath{m_a^2 \big|_{\rm FV}}}
\newcommand{\mafv}{\ensuremath{m_a \big|_{\rm FV}}}

\newcommand{\gagtv}{\ensuremath{g_{a\gamma} \big|_{\rm TV}}}
\newcommand{\gagfv}{\ensuremath{g_{a\gamma} \big|_{\rm FV}}}

\newcommand{\ttt}{\ensuremath{\tilde T_t}}
\newcommand{\tg}{\ensuremath{\tilde{g}_{\star}}}

\allowdisplaybreaks

\title{Axion Relic Pockets --- \\ a theory of dark matter }

\author[a]{Pierluca~Carenza~\orcidlink{0000-0002-8410-0345}}
\author[a]{Joshua~Eby~\orcidlink{0000-0003-0562-9177}}
\author[b,a]{Oksana Iarygina~\orcidlink{0000-0002-6222-1664}}
\author[a,1]{M.C.~David Marsh~\orcidlink{0000-0001-7271-4115}\note{Corresponding author.}}

\affiliation[a]{The Oskar Klein Centre, Department of Physics, Stockholm University, Stockholm 106 91, Sweden}
\affiliation[b]{Nordita, KTH Royal Institute of Technology and Stockholm University, Hannes Alfv\'ens v\"ag 12, 106 91 Stockholm, Sweden}

\emailAdd{david.marsh@fysik.su.se}

\abstract{We propose a new theory of dark matter based on axion physics and cosmological phase transitions. We show that theories in which a gauge coupling increases through a first-order phase transition naturally result in `axion relic pockets': regions of relic false vacua stabilised by the pressure from a kinematically trapped, hot axion gas.  
Axion relic pockets provide a
viable and highly economical theory of dark matter: the macroscopic properties of the pockets depend only on a single parameter (the phase transition temperature). 
We describe the formation, evolution  and present-day properties of axion relic pockets,  
and  outline how their phenomenology is distinct from existing dark matter paradigms. 
We briefly discuss how laboratory experiments and astronomical observations can be used to test the theory, and identify gamma-ray observations of magnetised, dark-matter-dense environments as particularly promising.
}

\begin{document} 
\maketitle
\flushbottom

\section{Introduction}
\label{sec:intro}
Gauge theories have two parameters: the coupling constant, $g$, and the theta-angle, $\theta$. Quantum gravity is expected to have no free parameters \cite{Green:1987mn, Polchinski:1998rr}, implying that $g$ and $\theta$ are, fundamentally, field-dependent. We consider a confining, hidden-sector gauge theory and refer to $\phi(x)=\Lambda/g^2(x)$ as the dilaton and $a(x)= \theta(x) f_a$ as the axion. The scales $\Lambda$ and $f_a$ (the `axion decay constant') are high-energy scales associated with the gauge theory. In this paper, we combine insights from cosmic phase transitions and axion physics to develop a new theory of dark matter. Our scenario is remarkably predictive and phenomenologically  distinct from existing dark matter paradigms.

Our starting point is the `hot Big Bang' era, when the energy density was dominated by the relativistic  plasma of the Standard Model. We assume  that axion particles are cosmologically present, at least in trace amounts, and that the dilaton undergoes a quantum phase transition from the `false vacuum', $\phi_{\rm FV}$, to the `true vacuum' satisfying $\phi_{\rm TV}< \phi_{\rm FV}$.  
The phase transition leads to an exponential increase in the axion mass, making the expanding true vacuum bubbles kinematically inaccessible to the axions. 
As the bubbles expand and seek to percolate,  axions are accelerated by repeated wall collisions and compressed into false vacuum regions that, for broad parameter ranges, eventually become 
stabilised by the hot axion gas. We refer to the resulting massive clumps as \emph{`axion relic pockets'} and show that they can easily comprise all of dark matter. The theory is highly economical and 
the predictions depend only on a few parameters. Depending on the time of the phase transitions, axion relic pockets vary in size from stellar to point-like, and in mass from  asteroid-like to intermediate particle physics scales. We discuss how laboratory experiments and astronomical observations may probe large parts of the parameter space, e.g.~by searching for highly characteristic signals based on axion-photon conversion of the hot gas. Across the  parameter space, a diverse set of experiments and observations may be devised to test the theory and potentially discover axion relic pockets.

\begin{figure*}[t!]  
\centering
\includegraphics[width= \textwidth]{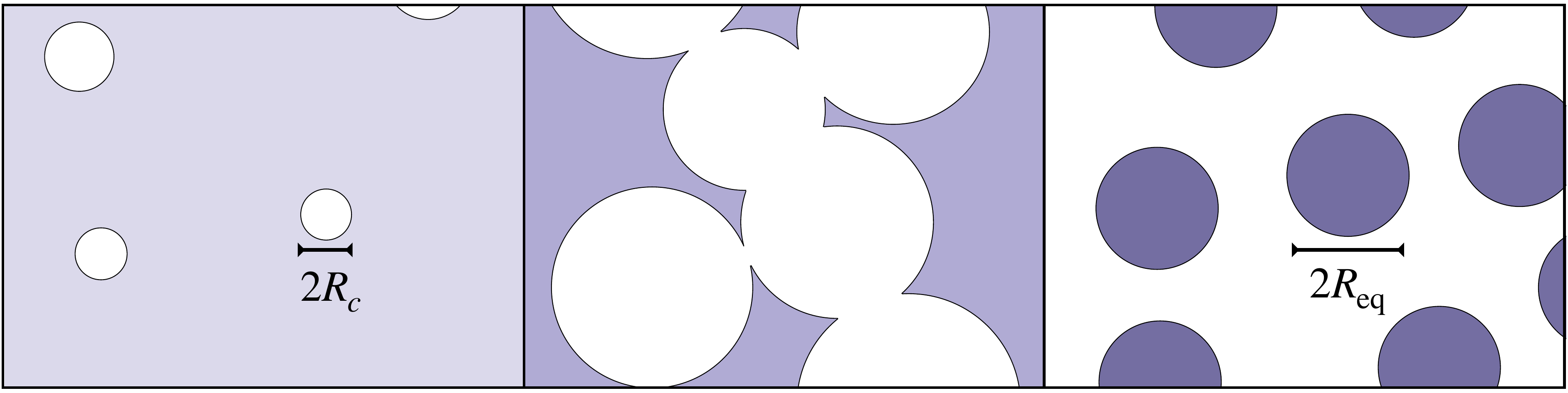}
\caption{Left: Bubbles of true vacuum with radii $R \gtrsim R_c$ nucleate through a dilaton phase transition in a background of axions (light blue). Middle: The expansion of the bubbles compress and accelerate the axions (blue), which are kinematically unable to enter the bubbles. Right: Axion relic pockets of radius $R_{\rm eq}$ form from regions of false vacuum stabilised by trapped, hot axion gas (dark blue). }
\label{fig:scenario}
\end{figure*}

Our theory builds on previous work on phase transitions and the fate of false vacua \cite{Coleman:1977py, Callan:1977pt}, spanning several decades. 
The QCD phase transition is now known to be a cross-over; however, early work considered the possibility of a first-order QCD phase transition and showed that `quark nuggets', i.e~pockets  of the high-energy phase could have formed, supported by the fermion degeneracy pressure of hypothetical strange quark matter \cite{Witten:1984rs}. In this context, 
and related most closely to our scenario,  
Ref.~\cite{Hindmarsh:1991ay} showed that if the QCD axion comprises dark matter (assumed to be produced from cosmic strings) and its mass increases by a moderate amount across the transition, axion pressure can contribute to, or even dominate, the internal pressure of nuggets. Related ideas involving axion domain walls \cite{Zhitnitsky:2002qa, Zhitnitsky:2021iwg} or hidden sector gauge theories \cite{Bai:2018dxf, Gross:2021qgx} have subsequently been developed. Transient `quark drops' or `droplets' of the high-temperature phase have been studied analytically and through hydrodynamical simulations in \cite{Rezzolla:1995br, Rezzolla:1995kv}, and more recently (and model-independently) in \cite{Cutting:2019zws, Cutting:2022zgd} (cf.~also \cite{Xie:2024mxr}). 

%
%
%

%

Considering thermal phase transitions of a hidden-sector Higgs field, references \cite{Kawana:2022lba} (cf.~also \cite{Gross:2021qgx, Lu:2022paj}), showed that `thermal balls' of the high-temperature phase stabilised by scalar radiation pressure may form, but, if stable, overclose the universe by many orders of magnitude and hence require some additional dilution mechanism to be consistent with observations. We note that these conclusions stem from the assumption of a sufficiently strong thermal phase transition, required for the trapping condition to be well-motivated in the setup of \cite{Kawana:2022lba}, and, as we show, do not carry over to dilaton phase transitions. Finally,  
%
%
Refs.~\cite{Baker:2021nyl,Kawana:2021tde} (see also \cite{Gehrman:2023qjn}) proposed collapsing relic pockets as seeds of primordial black holes (PBHs); however, it was subsequently shown that the presence of trapped relativistic particles 
decreases the compactness, making it more difficult for PBHs to form \cite{Lewicki:2022nba, Lewicki:2023ioy, Lewicki:2023mik}. 
Recently, the interplay of axions and phase transitions were elucidated in different contexts in 
\cite{Lee:2024oaz, Nakagawa:2022wwm}, corroborating that axion particles can become kinematically trapped. The relation between kinetic trapping and baryogenesis was discussed in \cite{Arakawa:2021wgz}.

The starting point of our work is the highly restrictive form of the dilaton-axion potential, reviewed in Sec.~\ref{sec:dilaton}, and the physics of tunneling in quantum field theory, reviewed in Sec.~\ref{sec:tunneling}. We present the physics of a dilaton phase transition in Sec.~\ref{sec:PT}, and the formation of axion relic pockets in Sec.~\ref{sec:pockets}. We calculate the present-day cosmological pocket abundance in Sec.~\ref{sec:dm} and the phenomenologically most relevant properties of axion relic pockets (e.g.~their mass, size and internal temperature) in Sec.~\ref{sec:pheno}. In Sec.~\ref{sec:obs}, we outline how experiments and observations may search for axion relic pockets in  different parts of the theory parameter space, before concluding in Sec.~\ref{sec:disc}.

%

\section{Dilatons \& Axions} 
\label{sec:dilaton}
The interactions of axions and dilatons are determined by non-perturbative physics, which generates an effective potential for the fields. The leading-order gauge theory instanton and anti-instanton contributions give a vacuum energy of the form 
\begin{align}
    V_{\rm inst.} &= M^4 \, e^{-S_{\rm inst.}} 
    = M^4 \, e^{-\frac{8 \pi^2}{g^2}} \Big(1 - \cos(\theta) \Big) \, .
    \label{eq:instanton}
    \end{align}
Throughout this paper, we consider non-perturbative effects of a hidden-sector confining  gauge theory. The scale $M$ is expected to be an ultraviolet energy scale (e.g.~the Planck scale, or intermediate between the Planck scale and the string scale \cite{Demirtas:2021gsq}).   
    
The dilaton and axion promote the field theory parameters $1/g^2$ and $\theta$ to dynamical fields, and the instanton potential governs their interactions:
    \begin{align}
   V_{\rm eff}(\phi, a) =M^4 \, e^{-8 \pi^2 \frac{\phi}{\Lambda}} \left(1 - \cos\Big(\frac{a}{f_a}\Big) \right)
    \, , \label{eq:Veff}
\end{align}
up to an irrelevant phase in the argument of the cosine.  Clearly, theories that are weakly coupled in the ultraviolet (UV) generate large instanton actions, e.g.~$S_{\rm inst}= 50$ for $\alpha= g^2/4\pi = 1/25$, as motivated in Grand Unified Theories. In string theory, exotic instantons, e.g.~Euclidean D3-branes, may also contribute to the effective potential, generating actions of the form $S_{\rm ED3} = T_3 {\rm Vol}(\Sigma)$, where $T_3$ is the brane tension and $\Sigma$ is a cycle volume of the compactification manifold. In these theories, the cycle's volume modulus controls the gauge theory coupling (large volume meaning weak coupling) and corresponds to our dilaton field. Studies of axions in string theory have revealed that large actions, $S \sim {\cal O}(150\text{--}200)$ \cite{Svrcek:2006yi} or even larger \cite{Demirtas:2018akl, Demirtas:2021gsq}, are ubiquitous across different compactification schemes. 

\begin{figure}
    \centering
    \includegraphics[width=0.4\linewidth]{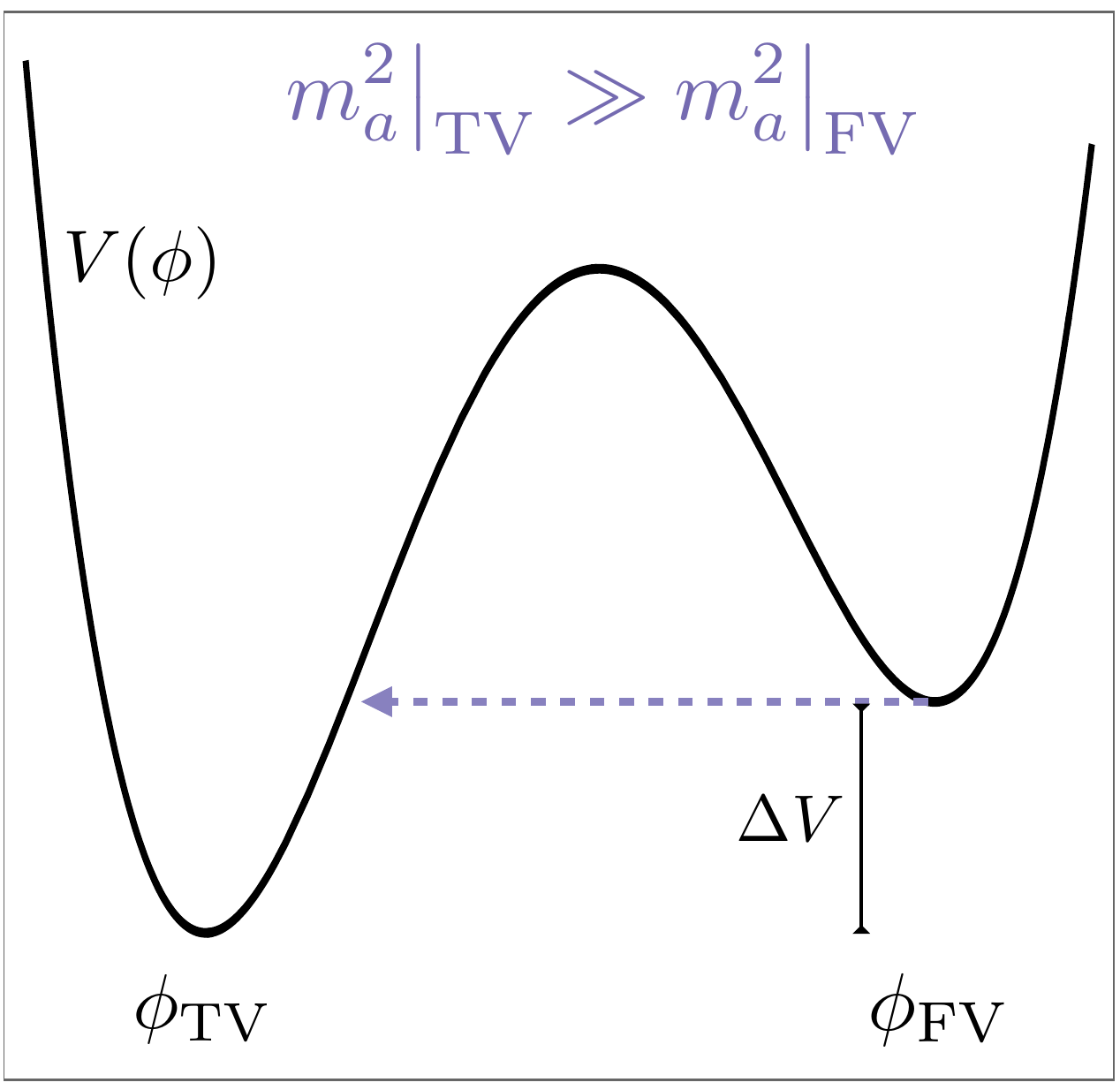}
    \caption{Schematic representation of the dilaton potential.}
    \label{fig:potential}
\end{figure}

The dilaton potential receives additional corrections, both non-perturbative and perturbative, in addition to that of Eq.~\eqref{eq:Veff} (cf.~\cite{McAllister:2023vgy} for a recent review).  
We will be agnostic about the detailed form of $V_{\rm tot}(\phi)$ and only assume that, in addition to the true vacuum at $\phi_{\rm TV}$, 
it features a metastable vacuum 
with $\phi_{\rm FV} > \phi_{\rm TV}$ so that the relevant region of the dilaton potential is of the schematic form of Fig.~\ref{fig:potential}.

\section{Cosmological Phase Transitions} 
\label{sec:tunneling}
Cosmological phase transitions are a generic feature of theories of physics Beyond the Standard Model. We consider the simplest scenario of a quantum, first-order phase transition resulting from the tunnelling of the dilaton from  $\phi_{\rm FV}$ to $\phi_{\rm TV}$, cf.~Fig.~\ref{fig:potential}. The dilaton is assumed to have negligible temperature, as is natural in theories with many weakly coupled hidden sectors. 
The true-vacuum nucleation rate per unit volume and time 
can be found from the `bounce' solution and
is given by \cite{Coleman:1977py}
\begin{align}
\frac{\Gamma}{\cal V} = \left( \frac{S_4}{2\pi}\right)^2 {\cal D}^2\, e^{-S_4} \, ,
\label{eq:Gamma}
\end{align}
where $S_4$ is the `bounce' action \cite{Coleman:1977py, Callan:1977pt},  and ${\cal D}$ denotes a 
(square-root) determinant of the quadratic fluctuation operator around the bounce 
\cite{Callan:1977pt, Baacke:2003uw}~(cf.~\cite{Ekstedt:2023sqc} for recently developed, efficient code for calculating tunnelling rates). The field profile of the bounce is $O(4)$ symmetric in Euclidean spacetime and simplifies significantly in the thin-wall limit, where the transition from $\phi_{\rm TV}$ inside the bubble to $\phi_{\rm FV}$ outside the bubble can be approximated as a sharp step at radius $R(t)$. In the absence of retarding forces,  expanding bubbles nucleate with radii above the critical radius, 
\begin{align}
 R_c = \frac{3\sigma}{\Delta V} \, ,
    \label{eq:Rc}
\end{align}
    and quickly accelerate to relativistic speeds \cite{Coleman:1977py}:
$
    \dot{R}(t) = t/\sqrt{R_c^2 + t^2}$.
The two relevant phase transition parameters are the wall tension, 
$
\sigma= \int_{\phi_{\rm TV}}^{\phi_{\rm FV}} d\phi \sqrt{2 V(\phi)} 
$  (cf.~\cite{Brown:2017cca} for a discussion) and the  potential energy difference, $\Delta V = V(\phi_{\rm FV})- V(\phi_{\rm TV})$,  
 cf.~Fig.~\ref{fig:potential}. As we consider a quantum, vacuum transition, $\Delta V$ is independent of the Standard Model temperature, $T$, and is equal to both the pressure difference and the  `latent heat' (defined for thermal transitions as $\Delta V - T_t \tfrac{d \Delta V}{dT}$, where $T_t$ is the temperature at the transition). To avoid confusion, we will refer to $\Delta V$ as the potential energy difference throughout this paper.  

The equation of motion of the (thin) spherical bubble wall follows from the Klein-Gordon equation and is given by \cite{Darme:2017wvu}
\begin{align}
    \ddot R + \frac{2(1 - \dot{R}^2)}{R} = {\mathfrak s}\frac{\Delta V \big(1 - \dot{R}^2 \big)^{3/2}}{\sigma}\left(1 - \frac{\cal P}{\Delta V}\right) \, ,
     \label{eq:EoMpocket}
\end{align}
where ${\mathfrak s}=1$ and ${\cal P}>0$ denotes the isotropic pressure exerted by the ambient axion background on the expanding bubble wall. 
Equation \eqref{eq:EoMpocket} also applies to a contracting, spherical false vacuum pocket by taking  $\mathfrak{s}=-1$, and letting ${\cal P}>0$ denote the outward pressure of the trapped axion gas
\cite{Gross:2021qgx,Kawana:2022lba, Lewicki:2023mik}.
  %


\section{The Dilaton Phase Transition}
\label{sec:PT}

We postulate a dilaton phase transition in which $\phi$ jumps from $\phi_{\rm FV}$ to  $\phi_{\rm TV}$ at time $t_t$ through bubble nucleation, cf.~Fig.~\ref{fig:potential}. The potential energy difference, $\Delta V$, is assumed to be sub-leading to the radiation energy density of the Standard Model plasma at the time of the transition, so that  $\rho_{\rm tot}(t_t) \approx \rho_{\rm rad} = \tfrac{g_\star}{30} T_t^4$, where  $T_t=T(t_t)$  is the temperature at the time of the transition and $g_\star(T)$ counts the number of relativistic species in equilibrium. At the transition time, one bubble nucleates per Hubble-volume and Hubble-time, so that, approximately, 
\begin{align}
    \Gamma/{\cal V} \approx H^4(t_t) = 
    \left(\frac{\pi^2 g_\star}{90} \right)^2 \left(\frac{T_t}{M_{\rm Pl}} \right)^4 T_t^4
    \, ,
    \label{eq:Tt}
\end{align}
where we have used the Friedmann equation in the last step and $M_{\rm Pl} = 2.4\cdot 10^{18}$ GeV denotes the reduced Planck mass.
In the absence of impeding forces, the dilaton bubbles rapidly grow, coalesce and eventually percolate to fill the entire universe with the true vacuum phase. 

An initial abundance of axion particles is guaranteed from the misalignment mechanism  if $\mafv \gtrsim 3H(t_t)$ \cite{Abbott:1982af, Preskill:1982cy, Dine:1982ah}, and may also be present through other production mechanisms, e.g.~as a relativistic `Cosmic Axion Background' formed from reheating or by the subsequent decay of heavy fields \cite{Conlon:2013isa, Cicoli:2012aq, Higaki:2012ar}. As we will show,  the properties of this population are largely unimportant in the simplest versions of our scenario, as long as the initial particle number density, $n_a$, is non-vanishing.  
%

\begin{figure}
    \centering
\includegraphics[width=0.7\linewidth]{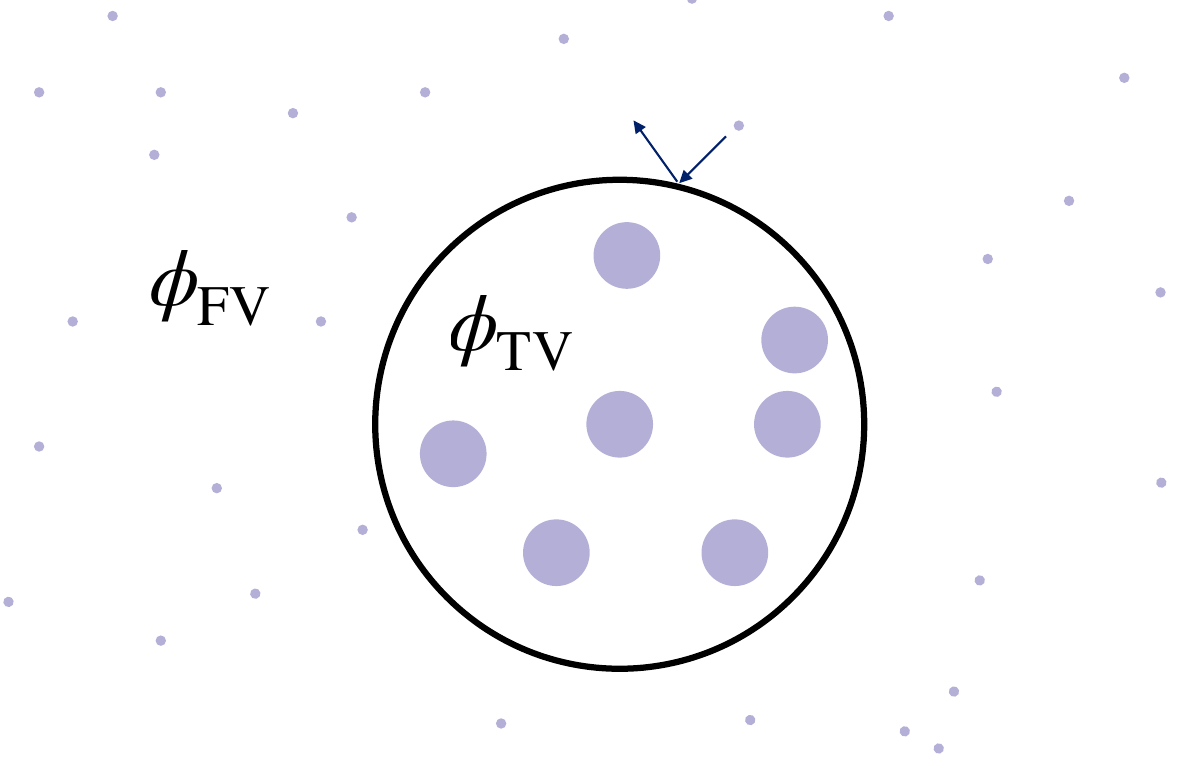}
    \caption{ The axions (purple dots) are lighter (illustrated as smaller) in the ambient false vacuum as compared to inside the true-vacuum bubbles. At nucleation, the energy released by a dilaton bubble of radius $R$ (i.e.~$\Delta V \cdot \tfrac{4\pi}{3} R^3$) is channeled into dilaton surface energy  ($\gamma_w \sigma \cdot 4\pi R^2$) and axion rest mass energy ($\Delta m_a\, n_a \cdot \tfrac{4\pi}{3} R^3$). Axions located outside the nucleated bubble can be kinematically forbidden from entering them, resulting in reflective boundary conditions for the axions on the dilaton bubble wall (illustrated by the arrows).}
    \label{fig:kinematics}
\end{figure}

A direct consequence of the restrictive form of the axion-dilaton potential, Eq.~\eqref{eq:Veff}, is that the axion mass grows exponentially from the false to the  true vacuum:
\begin{align}
    \masqtv = \masqfv \, {\rm exp}\Big[ S_{\rm FV}\, \tfrac{|\Delta \phi|}{\phi_{\rm FV}} \Big] \, ,
    \label{eq:Deltam}
\end{align}
where $\Delta \phi = \phi_{\rm TV} - \phi_{\rm FV}$ and $S_{\rm FV}$ is the false-vacuum value of the action of the non-perturbative effects generating the axion potential. Given the large instanton actions of weakly coupled theories (at high energies),  the axion mass can increase by many tens of orders of magnitude through the transition. 
%
%
The large increase in the axion mass has two consequences:
\begin{enumerate}
    \item {\bf Modified bubble energetics.}
The energy released by the nucleation of a true-vacuum bubble of radius $R$ is $\Delta V\cdot \tfrac{4\pi}{3} R^3$. In the absence of axions, this energy is converted to the surface energy of the bubble wall and, for $R> R_c$, the kinetic energy of the accelerated bubble growth. In the presence of axions located \emph{inside} the nucleated bubble, some of the released energy is transferred to the change in rest mass of the axion particles:  $\Delta m\, n_a \cdot \tfrac{4\pi}{3} R^3$.

For the bubble dynamics, this leads to an  `effective released energy', $\widetilde{\Delta V}(t)= \Delta V - \Delta m\, n_a(t)$, and an effective critical radius, $\widetilde R_c(t) =R_c/(1- \Delta m\, n_a/\Delta V)$, that both are time dependent. If the energy cost associated with the increasing axion masses is sufficiently large (which is model-dependent), the transition is blocked until $n_a(t) <\Delta V/\Delta m$. 
Our scenario is rather insensitive to whether such blocking has occurred, and for $R_c \ll R_H = 1/H(t_t)$, only a small fraction of the axions are located inside the nucleated bubbles. Even very weak, but model-dependent, couplings of these massive, true-vacuum axions to other fields can make them unstable so that they quickly decay. Conservatively, we will neglect their contribution to the dark matter density. 
\item {\bf Axion trapping.} For a sufficiently large $\masqtv$, axions  located outside the true-vacuum bubbles  
are kinematically forbidden from entering them,   cf.~Fig.~\ref{fig:kinematics}.
 The expanding dilaton bubble walls are, from the axions' perspective, perfectly reflective and rapidly expanding surfaces. Reflective boundary conditions are guaranteed for axions with a total energy less than $\matv$. Considering, for example, thermally distributed axions with temperature $T_a<\matv$, the transmission rate through the bubble wall is Boltzmann suppressed, and only the high-energy tail of the distribution is allowed to enter the bubble. In numerical simulations, 
 reflective boundary conditions have been found to provide an 
 excellent approximation as long as  $T_a \lesssim {\cal O}(10\, \matv)$ \cite{Hindmarsh:1991ay, Lewicki:2023mik}. Throughout this paper, we assume reflective boundary conditions for the axions, and we return to verify this assumption and discuss its implications in Sec.~\ref{sec:disc}.  We 
 now focus on an initial population of non-relativistic axions and discuss the cosmological impact of their trapping. 
\end{enumerate}

As the nucleated bubbles expand, they plough through the background of non-relativistic axions that are kinematically forbidden from entering them.  The exerted pressure from the axion gas on the bubble wall is given by (cf.~Appendix \ref{app:pressure}),
\begin{align}
    {\cal P} = \frac{\gamma_w^2}{6\pi^2} \int_0^\infty dp\, p^2 f(p) \frac{(p-E v_w)^3}{p E} \approx \frac{\delta^2  \gamma_w^2 |v_w|^3}{12 \pi^2} m_a^4\big|_{\rm FV} \, , 
      \hspace{1.8cm} \rlap{\text{[non-rel]} } \hphantom{000}
     \label{eq:Pnonrel}
\end{align}
where we have assumed a homogeneous and isotropic axion distribution function, $f(p)$, and denoted the axion energy by $E$, and the wall speed and gamma factor in the plasma frame  respectively by $v_w$ and $\gamma_w$. In the last step, we have, for concreteness and conservatively, specialised to a uniform distribution with $p$ in the range $0 \leq p \leq \delta\cdot \mafv$, with $\delta \ll1$. From Eq.~\eqref{eq:EoMpocket}, the terminal velocity of the bubble wall satisfies ${\cal P}/\Delta V=1$.  For non-relativistic axions with $m_a^4\big|_{\rm FV} \ll \Delta V$ (which we will see is the most relevant case), the pressure exerted on the bubble wall can be neglected if $\gamma^2_w \lesssim \Delta V/(\delta^2 m_a^4\big|_{\rm FV})$; this inequality  typically holds until bubble coalescence. Consequently, non-relativistic axions do not impede the bubbles, which `run away' with gamma-factors given by the bounce solution, $\gamma_w = R/R_c$, and do not reach a terminal velocity before nearly colliding with other bubble walls.\footnote{Note that previous work on relic false vacuum remnants assumed non-relativistic terminal velocities, as is well-motivated in a thermal phase transition context, cf.~e.g.~\cite{Witten:1984rs, Hindmarsh:1991ay, Bai:2018dxf, Gross:2021qgx, Kawana:2022lba}. } The advancing bubble walls are led by  increasingly dense  shock fronts of  reflected axions.



The phase transition is not instantaneous. At $t_t$, the  distance between bubble nucleation sites is approximately $d(t_t)\approx R_H
=2 t_t$. This distance grows with the Hubble expansion like  $d(t) = d(t_t) \frac{\mathfrak{a}(t)}{\mathfrak{a}(t_t)} = 2 \sqrt{t_t t}$ where $\mathfrak{a}$ denotes the Friedmann-Robertson-Walker scale factor. 
The coalescence time, $t_{\rm coll}$, 
can be found from the equation
$$
\frac{d(t_{\rm coll})}{2} = \sqrt{R_c^2 + (t_{\rm coll}-t_t)^2} \, ,
$$
which gives
\begin{align}
    t_{\rm coll} \approx
\frac{t_t}{2}
\left(  
3 + \sqrt{5 - 4 R_c^2/t_t^2}
\right) \approx 2.6 t_t \, ,
\label{eq:tco}
\end{align}
where the last step applies to $R_c/R_H \ll 1$. In this limit, $\mathfrak{a}(t_{\rm coll})/\mathfrak{a}(t_t) \approx 1.6$. 
Following coalescence, the relativistic axion fronts strike  multiple bubble walls, and begin to exert appreciable pressure. For $t>t_{\rm coll}$, we change the perspective from expanding bubbles to contracting false vacuum pockets filled by axion gas, i.e.~axion relic pockets.


\begin{figure}[t]
    \centering
  \includegraphics[width=\linewidth]{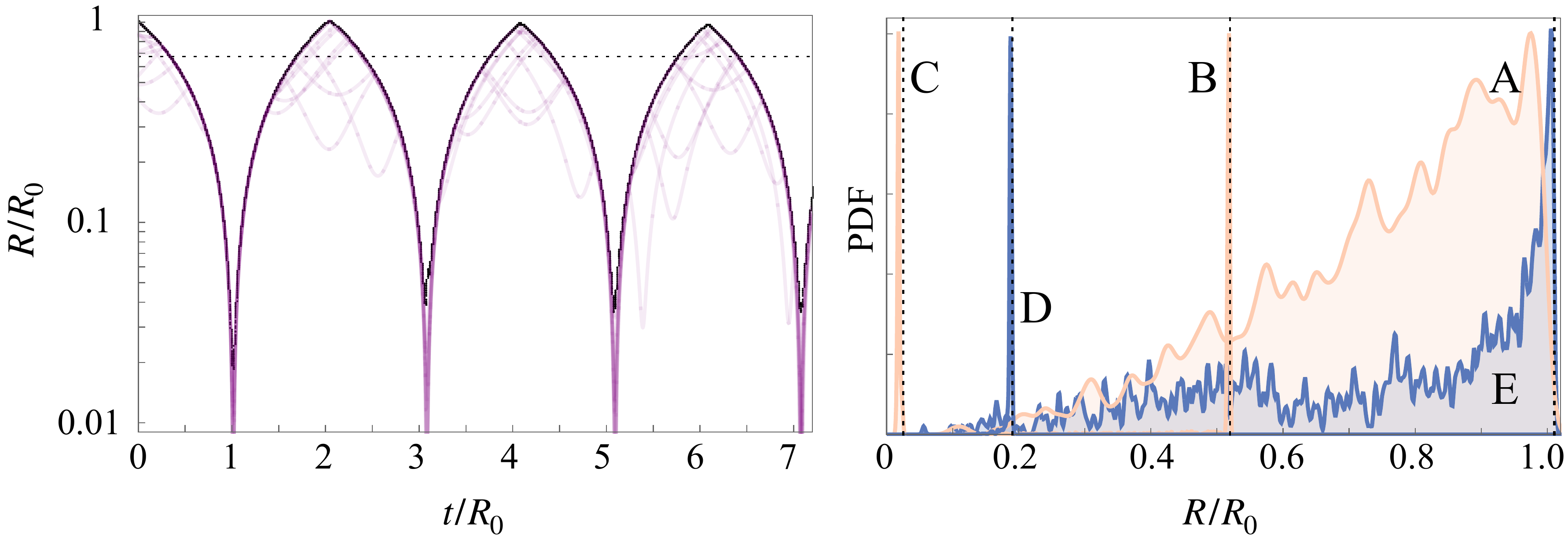} 
    \caption{Left: The evolution of the spherical pocket wall (black) together with the trajectories of 10 (out of  $10^3$) randomly selected axion trajectories (purple). Right: radial axion distributions (Probability Distribution Functions; rescaled to fit on plot) for the contracting pocket (pale orange) at (A)   $t/R_0=0$, (B) $t/R_0=0.5$ and (C) $t/R_0=1.0$, and for expanding pockets (blue) at (D) $t/R_0=1.2$ and (E) $t/R_0=2.0$. Adjacent vertical dotted lines indicate the corresponding wall position (coincidental for $t/r_0 =0$ and $2$).   
    Here $\dot R(0)=-0.999$, $R_c/R_0 =0.01$, which corresponds to $\alpha=0.22$. The expected equilibrium radius is $R_{\rm eq}/R_0 = 0.67$, as indicated by the dashed line in the left panel.  }
    \label{fig:pocket_sim}
\end{figure}

\section{Axion Relic Pockets}
\label{sec:pockets}
We now discuss the dynamics of relic pocket formation and derive their key properties, in particular their typical radius and mass. We then, in Sec.~\ref{sec:dm}, return to the cosmology of axion relic pockets.

The phase transition cannot percolate, since axions trapped in the relic pockets cannot escape the false vacuum. In this section, we discuss the dynamics of a single pocket in isolation, and,   for simplicity, we restrict to spherical pockets.\footnote{The sphericalisation process of the colliding bubble walls is an interesting topic for  numerical investigation that goes beyond the scope of this work.} We take the initial radius to be  $R_0= 1/H(t_{\rm coll})$, and the initial gamma-factor to satisfy $\gamma_w \gg 1$. The initial axion gas pressure  is assumed to be subleading to $\Delta V$,  ${\cal P}(t_{\rm coll})\ll \Delta V$.   In Appendix \ref{sec:ODE}, we derive a set of ordinary differential equations governing the coupled evolution of the pocket wall and the gas pressure, assuming a homogeneous and isotropic axion distribution function, as is common in the literature (in which rapid thermalisation often is assumed). However, we also show that such distribution functions are incapable of capturing the rapidly increasing pressure in the axion gas, and lead to violations of causality. Thus, this approach is not suitable for the study of axion relic pockets.

Instead, we follow the approach of   \cite{Lewicki:2023mik} and analyse the contracting system through $N$-body simulations with a finite number of ultrarelativistic axion particles subject to reflective boundary conditions at the wall. Figure \ref{fig:pocket_sim} shows an example of  the result of such a simulation with 1000 particles,  $R_c/R_0 = 0.01$ and an initial wall velocity $\dot R = -0.999$ ($\gamma_w =22.4$). The axion pressure is initially $\ll \Delta V$, and the initial axion energy fraction $E^{\rm tot}_{a}/E_{\rm tot} \approx 0.21$. Axions reflected by the contracting wall form a radially localised, spherical front in advance of the wall (cf.~the right panel of Fig.~\ref{fig:pocket_sim}). Shortly before the wall would collapse, the front `passes through the origin' and strikes the contracting dilaton wall, causing it to bounce back with a high $\gamma_w$ factor. The expanding pocket is trailed by a dense axion front that counteracts the deceleration of the wall caused by $\Delta V$ (and to a lesser extent the surface tension), a mechanism that becomes inefficient only when $R\approx R_0$, after which the pocket again contracts and the cycle repeats. 
%

Our simulations remain stable over multiple cycles, but do not unambigously indicate a timescale for the damping of the oscillations (the same holds true for the data from the most similar simulations of \cite{Lewicki:2023mik}, which we have analysed). Quartic axion interactions are expected to lead to momentum dispersion and damping of the oscillations at a time 
scale of $\tau \sim 64\pi^2 \mathcal{G} E_s^3/(\rho_s \lambda^2)$, where $\rho_s$ and $E_s$ are the energy density and characteristic energy per particle in the propagating shocks, $\lambda$ is the dimensionless quartic self-coupling, and $\mathcal{G}$ is a geometric factor (of order $(R_c/R_0)^2$ for a spherical pocket). 
It is possible that other processes could lead to a faster damping.

When the oscillations subside, the relic pockets stabilise at a finite equilibrium radius, which from pressure equality (cf.~Eq.~\eqref{eq:EoMpocket}) is given by 
\begin{align}
    R_{\rm eq} = \frac{2 \sigma}{{\cal P} - \Delta V} = \frac{2 }{3} \frac{R_c}{{\cal P}/\Delta V -1}\, .
\end{align}
From our simulations we find that the pockets stabilise at $R_{\rm eq} \sim R_0 \gg R_c$, implying ${\cal P}/\Delta V -1\ll1$ in equilibrium. One can understand this as a consequence of energy conservation. 
The mass of the equilibrium axion relic pocket is given by 
\begin{align}
    M_{\rm pocket} &= \sigma {\cal A}_{\rm eq} + \left(\Delta V + 3 {\cal P} \right) {\cal V}_{\rm eq} 
    = \Delta V\, {\cal V}_{\rm eq} \left(4 + \frac{R_c}{R_{\rm eq}} \right) \approx 4 \Delta V {\cal V}_{\rm eq} \, , 
    \label{eq:Mpocket1}
\end{align}
where ${\cal A}_{\rm eq}=4\pi R_{\rm eq}^2,~{\cal V}_{\rm eq}=\frac{4\pi}{3} R_{\rm eq}^3$, $\rho_{\rm gas} = 3{\cal P}$, and the last step applies to $R_{\rm eq} \gg R_c$. Thus, the mass of large pockets consists of 1/4 vacuum energy and 3/4 relativistic axion energy. Moreover, the mass of the equilibrium pocket must be equal to the total energy of the contracting axion-wall system, which initially, at $t=t_{\rm coll}$, is distributed as
\begin{align}
E_{\rm tot} = \gamma_{w}(t_{\rm coll}) \sigma {\cal A}_0 + \Delta V {\cal V}_0 + E_{\rm gas}^{(0)} \, .
\end{align}
Here ${\cal A}_0$ and ${\cal V}_0$ respectively denote the area and volume of a sphere with radius $R_0$, and we parametrise the initial gamma-factor of the wall as $\gamma_{w}(t_{\rm coll}) = \alpha R_0/R_c$, where we expect $0 < \alpha \lesssim 1.6$, cf.~Eq.~\eqref{eq:tco}. 
As above,  the axion gas energy is initially small, 
$E_{\rm gas}^{(0)} \ll E_{\rm tot}$, and can be neglected. This gives another equation for the mass of the pocket:
\begin{align}
    M_{\rm pocket}= (1 + \alpha) \Delta V {\cal V}_0 \, .
    \label{eq:Mpocket2}
\end{align}
The $R_{\rm eq}/R_c\gg1$ limit of  Eqs.~\eqref{eq:Mpocket1} and \eqref{eq:Mpocket2} then consistently gives
\begin{align}
 R_{\rm eq} = \left(\frac{1+\alpha}{4}\right)^{1/3}R_0
\label{eq:Req}
\end{align}
corresponding to ${\cal P}/\Delta V -1 =\tfrac{2}{3} \tfrac{R_c}{R_0} \Big(\frac{4}{1+\alpha}\Big)^{1/3} \ll 1$. Thus, axion relic pockets have radii comparable to the Hubble radius at the time of their formation.

\section{Dark Matter Cosmology}
\label{sec:dm}
We now return to the cosmology of the dilaton phase transition with the goal of calculating the present-day abundance of axion relic pockets. The  pocket number density   dilutes like $\sim 1/{\mathfrak a}^3$ with cosmic expansion. The total  energy density of the population of axion relic pockets redshifts like dust provided their peculiar velocities are sufficiently small. However, since they originate from the near-collision of multiple relativistic dilaton bubbles, one may expect them to form with large peculiar velocities, and to subsequently be slowed by Hubble expansion between $t_{\rm coll}$ and some final time, $t_f$. Reference \cite{Cutting:2022zgd}  studied hydrodynamical  `droplets' observed in large-scale simulations and demonstrated that the peculiar velocities can be heavily suppressed relative the wall velocities. Determining the velocity distribution of axion relic pockets at $t_{\rm coll}$ requires dedicated, large-scale simulations, which goes beyond the scope of this work. Here, we instead model  the dilaton-axion energy density as follows:
\begin{align}
    \rho(\phi , a) \sim 
    \begin{cases}
        {\rm const.} & \text{for } t\leq t_t \, ,\\ 
        \mathfrak{a}^{-4} & \text{for } t_t<t \leq t_f \, , \\
               \mathfrak{a}^{-3} & \text{for } t_f\leq t \, .
    \end{cases}
    \label{eq:rhocases}
\end{align}
Before $t_t$, the total energy density available for the phase transition is $\Delta V$. This energy density is first transferred to the formation and acceleration of the bubble walls and, after $t_{\rm coll} \in (t_t,\, t_f)$, the heating of the axion gas and motion of the pockets. In Eq.~\eqref{eq:rhocases}, we have assumed that the energy redshifts like radiation during between $t_t$ and $t_f$.
We expect this to be
conservative for $t_t<t<t_{\rm coll}$ when the energy density is shared between relativistically expanding bubbles and yet-untouched false-vacuum regions, in the sense that the true energy density likely redshifts somewhat slower. For $t_{\rm coll}<t<t_f$, the energy density is in the form of rapidly moving axion relic pockets, and we expect the $\sim \mathfrak{a}^{-4}$ scaling to be accurate.

With these assumptions,  the energy density left to the pockets at $t_f$ is given by $\Delta V \epsilon^4$, where 
\begin{align}
    \epsilon = \left(\frac{\mathfrak{a}(t_t)}{{\mathfrak a}(t_f)}\right) = \left(\frac{t_t}{t_f}\right)^{1/2} \, .
    \label{eq:epsilon}
    \end{align}
    We can bound this parameter as $\epsilon^4 < (t_{t}/t_{\rm coll})^2 \approx 0.15$ using Eq.~\eqref{eq:tco}; however, we expect phase transitions where $R_c\ll R_H$ to produce values of $\epsilon$  well below this bound. 

At $t_f$, the axion-dilaton sector takes the form of a population of slow-moving axion relic pockets with number density $n_{\rm pocket}(t_f)$ and energy density  
%
%
\begin{align}
\rho_{\rm pocket}(t_f)=\epsilon^4  \Delta V = n_{\rm pocket}(t_f) M_{\rm pocket} \, ,
\label{eq:energycons}
\end{align}
where  $M_{\rm pocket}$ is given by Eq.~\eqref{eq:Mpocket1}. 
The filling fraction of relic pockets at $t_f$ is then given by
\begin{align}
    f(t_f) = {\cal V}_{\rm eq} n_{\rm pocket}(t_f) = \frac{\epsilon^4}{4 +  R_c/R_{\rm eq}} \approx \frac{\epsilon^4}{4}
\, ,  
\end{align}
where in the last step we have restricted to $R_{\rm eq}\gg R_c$.

\begin{figure}[t]
    \centering
    \includegraphics[width=0.8\linewidth]{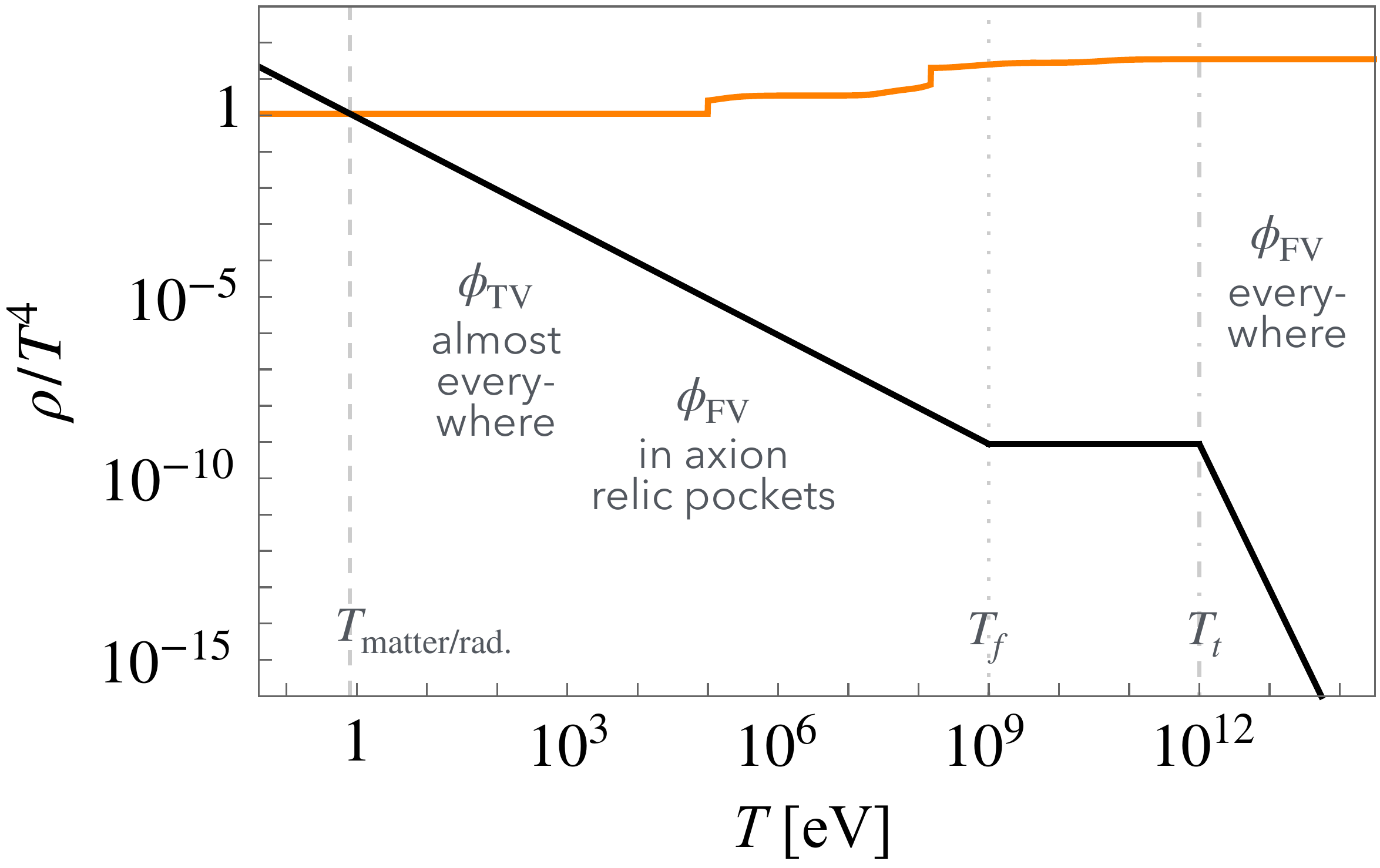}
    \caption{The energy density of the Standard Model plasma (orange) and that of the dilaton-axion system (black). For $T>T_t$, the dilaton energy density is constant (and equal to $\Delta V$), for $T_t > T > T_f$,  it is expected to red-shift approximately like radiation, and for $T<T_f$, it redshifts like matter and can overtake the radiation density at matter-radiation equality ($T_{\rm matter/rad} \approx0.8\, {\rm eV}$). In this example, $T_t= 1\, {\rm TeV}$, $T_f=1\, {\rm GeV}$ (corresponding to $\epsilon=10^{-3}$), $\alpha=1.6$, and only Standard Model particles contribute to $g_\star(T)$.}
    \label{fig:a4rho}
\end{figure}

The energy density of axion relic pockets redshifts like matter for $t\geq t_f$ and eventually overtakes the Standard Model plasma as the universe's leading energy source, cf.~Fig.~\ref{fig:a4rho}. 
The present-day energy density of axion relic pockets (at $t=t_0$) is 
\begin{equation}
    \rho_{\rm pocket}(t_0) = \epsilon^4 \rho_{\rm pocket}(t_t)  \left( \frac{{\mathfrak a}(t_f)}{{\mathfrak a}(t_0)} \right)^3 
    = \epsilon \Delta V \frac{g_{\star S}(T_0) T_0^3}{g_{\star S}(T_t) T_t^3} \, ,
\end{equation}
where, in the last step, we have used Eq.~\eqref{eq:epsilon} and entropy conservation. Here, $T_0$ denotes the Standard Model temperature at time $t_0$ and $g_{\star S}$ is the effective number of relativistic degrees of freedom in entropy. Axion relic pockets comprise all of the dark matter if
$
    \Omega_{\rm pocket} h^2 = 0.12
$, i.e.~if
\begin{align}
    \Delta V \big|_{\rm DM} &= \frac{0.12}{\epsilon}\frac{\rho_{\rm crit}}{h^2} \frac{g_{\star S}(T_t)\, T_t^3} {g_{\star S}(T_0)\, T_0^3} 
    = 19.5\;  {\rm GeV}^
4 ~
   \epsilon^{-1} \tilde{g}_{\star S}(t_t)
     \ttt^3 \, \, ,
\label{eq:DeltaVDM}
\end{align}  
where we
have used $\rho_{\rm crit}/h^2 = 8.1\cdot 10^{-47}\, {\rm GeV}^4$, $T_0=2.4\cdot 10^{-13}\, {\rm GeV}$ and $g_{\star S}(T_0)=3.94$.
For ease of notation, we have introduced $\ttt = T_t/{\rm TeV}$ and $\tilde{g}_{\star S}(t_t)= g_{\star S}(t_t)/106.75$.
As will become clear, the most interesting phase transitions occur before $t_t\approx1$ second, so we take $\tilde{g}_{\star S}(t_t)=\tilde{g}_{\star}(t_t)$ in what follows.


Fixing $\Delta V$ by Eq.~\eqref{eq:DeltaVDM} determines one of the two phase-transitions parameters, leaving a one-parameter family of solutions parametrised by $\sigma$ (or, equivalently, $R_c$  or $T_t$ using  Eqs.~\eqref{eq:Gamma} and \eqref{eq:Tt}). Note that the undetermined factor $\epsilon$ is not a free parameter of the theory, but can be uniquely fixed from simulations (for fixed $\Delta V$ and $\sigma$).
We note that for transitions completing well before matter-radiation equality, $\Delta V$ is highly suppressed relative to $\rho_{\rm rad}(t_t)$:
\begin{align}
\frac{\Delta V \big|_{\rm DM}}{\rho_{\rm rad}(t_t)} =   
5.6\cdot 10^{-11}\, 
\epsilon^{-1}\, \ttt^{-1} 
\, ,
\label{eq:energyfraction}
\end{align}
where we have neglected variations in $g_\star$ between $t_t$ and $t_f$. 

The transition temperature is the only free parameter of the simplest version of this theory and, a priori, it spans a remarkably large range of 25 orders of magnitude: $1\, {\rm eV} \lesssim T_t \lesssim 10^{16}\, {\rm GeV}$. Here, the lower limit  is set by the temperature of matter-radiation equality and the upper limit is set by the maximal energy scale of inflation \cite{Tristram:2021tvh}. However, we will see  that theoretical consistency considerations and observational constraints narrow down the range of permissable transition temperatures. Still, the broad range of possible transition temperatures results in a qualitative diversity in the properties of axion relic pockets, as we now discuss.

\section{Axion Relic Pocket Phenomenology}
\label{sec:pheno}
We are now ready to calculate the key properties of axion relic pockets, conditioned on these comprising all of dark matter, cf.~Eq.~\eqref{eq:DeltaVDM}.
The pocket mass is given by Eq.~\eqref{eq:Mpocket1} or, equivalently, Eq.~\eqref{eq:Mpocket2}. The Hubble radius at the time of the transition is given by
\begin{align}
    R_H(t_t) &= \frac{1}{H(t_t)} = \left(\frac{90 M_{\rm Pl}^2}{g_\star(t_t) \pi^2 T_t^4}\right)^{1/2} = 
    7.0 \cdot 10^{11}\, {\rm GeV}^{-1}~\tg^{-1/2}\, \ttt^{-2} \nonumber \\
    &=1.4 \cdot 10^{-4}\, {\rm m}~\tg^{-1/2}\, \ttt^{-2} \, .
\end{align}
Since $R_H \sim t$, we have that 
\begin{align}
    R_0 \approx R_H(t_{\rm coll}) = \frac{t_{\rm coll}}{t_t} R_H(t_t) =3.6 \cdot 10^{-4}\, {\rm m}\, ~\tg^{-1/2}\, \ttt^{-2}  \, ,
\end{align}
where, in the last step, we have used Eq.~\eqref{eq:tco}. The equilibrium radius  is given by Eq.~\eqref{eq:Req}, 
\begin{align}
 R_{\rm eq} =  1.1\cdot10^{12}\, {\rm GeV}^{-1} ~(1+\alpha)^{1/3}\, \tg^{-1/2}\, \ttt^{-2} = 2.3\cdot 10^{-4}\, {\rm m} ~(1+\alpha)^{1/3}\, \tg^{-1/2}\, \ttt^{-2}  \, .        
\end{align}
The characteristic relic pocket mass is then given by 
\begin{align}
    M_{\rm pocket} &= (1+\alpha) \frac{4\pi}{3} R_0^3 \Delta V\big|_{\rm DM} = 4.9 \cdot 10^{38}\, {\rm GeV} \left(\frac{1+\alpha}{\epsilon}\right)
    ~\tg^{-1/2}\, \ttt^{-3} 
     \nonumber \\
    &= 4.3 \cdot 10^{-19} M_\odot \left(\frac{1+\alpha}{\epsilon}\right)
     ~\tg^{-1/2}\, \ttt^{-3} \, .
\end{align}
Clearly, later phase transitions produce larger and heavier pockets. Large initial pocket peculiar velocities (i.e.~a small $\epsilon$) increase the pocket mass, but leave radii unchanged.  For all relevant parameter values, the stable pockets are much larger than their  Schwarzschild radius ($R_S=2GM$, where $G$ is the gravitational constant):
\begin{align}
\frac{R_{\rm eq}}{R_S} = 1.7\cdot 10^{11}  ~
    \frac{\epsilon\, \ttt }{(1+\alpha)^{2/3}}   
    \gg 1\, ,
\end{align}
which is consistent with the findings of \cite{Lewicki:2023mik}, which established a similar result also for the maximum compactness in the pocket oscillating phase. 

Assuming a local dark matter density of $\rho_{\rm dm} = 0.4\, \text{GeV/cm}^3$, the  average separation of axion relic pockets  is
\begin{align}
      D_{\rm pocket}(t_0) = \left(\frac{M_{\rm pocket}}{\rho_{\rm dm}}\right)^{1/3} =1.1\cdot 10^{11}~{\rm m}\, \left(\frac{1+\alpha}{\epsilon}\right)^{1/3}
    ~\tg^{-1/6}\, \ttt^{-1} \, . 
    \label{eq:Dt0}
\end{align}
The present-day pocket number density is $n_{\rm pocket}=D^{-3}_{\rm pocket}(t_0)$.
The rate at which a pocket will encounter an object of cross-sectional area $\sigma$ is given by
\begin{align}
    \Gamma_{\rm encounter} 
    \sim n_{\rm pocket} \sigma v_{\rm pocket} 
    = 10^{-6}\,{\rm year}^{-1} 
    \left(\frac{R}{R_\oplus}\right)^{2} 
    \frac{\tg^{1/2}\epsilon}{1+\alpha} \ttt^{3}
    \,,
    \label{eq:encounter}
\end{align}
where $R_\oplus$ is the radius of the Earth, we took the speed of the pocket $v_{\rm pocket}\simeq 10^{-3}$ (appropriate for virialized dark matter in the solar neighborhood), and we assumed a spherical body with $\sigma= \pi R^2$. The Earth would encounter one pocket per second for $T_t \gtrsim 3\cdot10^8\,{\rm GeV}$, and a detector of radius $1$~meter would encounter one pocket per second for $T_t \gtrsim 10^{12}\,{\rm GeV}$.

The hot axion gas comprises 3/4 of the energy of each pocket and is of great phenomenological interest. In the simplest scenario, the axion gas thermalises, e.g.~through self-interactions, before the present day ($t_0$), which makes the predictions independent of the initial abundance of axions at $t\leq t_t$. In this case, the axion temperature is given by 
\begin{align}
  T_{\rm eq} = \left(\frac{30}{\pi^2} \frac{\frac{3}{4} M_{\rm pocket}}{{\cal V}_{\rm eq}} \right)^{1/4} = 5.3~{\rm GeV}~ \frac{\tg^{1/4}}{\epsilon^{1/4}} \, 
      \ttt^{3/4}\, .
    \hspace{1.8cm} \rlap{\text{[thermal]} } \hphantom{000}
    \label{eq:Tthermal}
\end{align}
The number of axions contained in a typical pocket is given by 
\begin{align}
   N_{a\, \text{pocket}} &= n_a(T_{\rm eq}) \frac{4\pi}{3} R_{\rm eq}^3 = 1.0\cdot 10^{38}~\frac{(1+\alpha)^{1/2}}{\tg^{3/4} \epsilon^{3/4}}  \ttt^{-15/4} \, .
    \quad \rlap{\text{[thermal]} } \hphantom{000}
\end{align}
For theoretical consistency, the number of axions in each pocket must be sufficiently large to justify the assumption of sphericity and avoid pocket collapse, suggesting that  $N_{a\, \text{pocket}}\gg 1$. 
However, we note that, depending on the initial conditions, our numerical $N$-body simulations described in Sec.~\ref{sec:pockets}  avoid pocket collapse for quite low particle numbers, $N_{a\, \text{pocket}} \sim {\cal O}(10\text{--}1000)$. Most conservatively, the strict upper bound of at least having one axion per pocket implies (for a thermalised axion gas) $T_t \lesssim 10^{13}$ GeV, which removes around three orders of magnitude from the naive parameter space described in Sec.~\ref{sec:dm}.

If the axions have not yet thermalised, 
their characteristic energy is given by
\begin{align}
    \bar E_a =\frac{3}{4} \frac{M_{\rm pocket}}{N_{a\, \text{pocket}}} \,,
       \rlap{\hspace{3.7cm} \text{[non-thermal]} } \hphantom{000000000000000}
\end{align}
where $N_{a\, \text{pocket}}$ denotes the number of axions in each pocket which, if axion-number changing processes are inefficient,  is approximately given by the number of axions per Hubble volume at the time of the transition:
\begin{align}
    N_{a\, \text{pocket}} = n_a(t_t) \frac{4\pi}{3} R_H(t_t)^3 \, . \rlap{\hspace{2.8cm}\text{[non-thermal]}}
    \hphantom{000000000000000}
\end{align}
The factor $n_a(t_t)$ can be calculated given additional model assumptions. For example, if the initial axion population originates from the misalignment mechanism, the number density is given by
\begin{align}
    n_a(t_t) = \frac{1}{2} f_a^2 \mafv \theta_i^2 \frac{g_{\star S}(T_t) T_t^3}{g_{\star S}(T_{\rm osc}) T_{\rm osc}^3} \, , \rlap{\hspace{2cm}\text{[misalignment]}}
    \hphantom{000000000000000}
\end{align}
where $\theta_i$ denotes the initial misalignment angle and $T_{\rm osc}$ denotes the temperature when the axion field started to oscillate, which we relate to the mass through the approximate equality $\mafv \approx3 H(T_{\rm osc})$.
In this case, $N_{a\, \text{pocket}} \sim (f_a/m_a)^2 \gg1$ for phase transitions with $T_{\rm osc} \approx T_t$, and, in contrast to the thermal case, there is no reduction of the upper bound of $T_t$ from the inflationary scale $T_t \lesssim 10^{16}\, \text{GeV}$.
Setting $g_\star(T_t)=g_\star(T_{\rm osc}) =106.75$, we find the characteristic energy 
\begin{align}
    \bar E_a = 1.3\, {\rm keV}\, \frac{1}{\tilde f_{10}^2 \theta_i^2 } \frac{1+\alpha}{\epsilon}\, \tilde T_{\rm osc}\, ,
   \rlap{\hspace{2.6cm}\text{[non-thermal]}}
   \hphantom{000000000000000}
   \label{eq:nonthermal}
\end{align}
where $\tilde f_{10}=f_a/(10^{10}\, {\rm GeV})$ and $\tilde T_{\rm osc} = T_{\rm osc}/(1\, {\rm TeV})$. Interestingly, $\bar E_a$ is independent of the transition temperature. The relation between \mafv~and 
the oscillation temperature is 
$
    \mafv=4.3\, {\rm meV}\, \tilde g_\star(T_{\rm osc}) \tilde T_{\rm osc}^2
    $.\footnote{For many consistent values of $\matv$, misalignment axion can result in `blocked' phase transitions of then type mentioned in Sec.~\ref{sec:PT}. For such transitions, we expect $R_c/R_0$ to be non-negligible, resulting in a delayed phase transition, comparatively larger critical bubbles, and  slower wall velocities. Moreover, the impact of heavy axions inside true vacuum bubbles may not be negligible. We expect to return to this scenario in future work.}

In sum, in the simplest scenario of a  thermalised axion gas, the size, mass and temperature depend only on a single parameter: $T_t$  (or, equivalently, $\sigma$). For a non-thermal gas, the characteristic energy is, in general, more model dependent as it is inversely proportional to the initial axion number density  at $t_t$ (the mass and radius depend only on the pressure and are unaffected by this model dependence). Surprisingly however, in the well-motivated case of an initial population of misalignment axions that are too weakly coupled to thermalise, 
the characteristic energy is \emph{independent of $T_t$}, and is most strongly dependent on  the axion parameters $\mafv$, $f_a$ and $\theta_i$. We emphasise that, even in the non-thermal case, the dark matter abundance is determined by the dilaton parameters (in particular, $\Delta V$, cf.~Eq.~\eqref{eq:DeltaVDM}), and many axion models that would not be able to comprise all of the dark matter through the standard misalignment mechanism (or cosmic strings) can provide the gas pressure of axion relic pockets, and, in our scenario, be responsible for most of the energy in dark matter. 

\begin{figure}[t]
    \centering
    \includegraphics[width=0.9\linewidth]{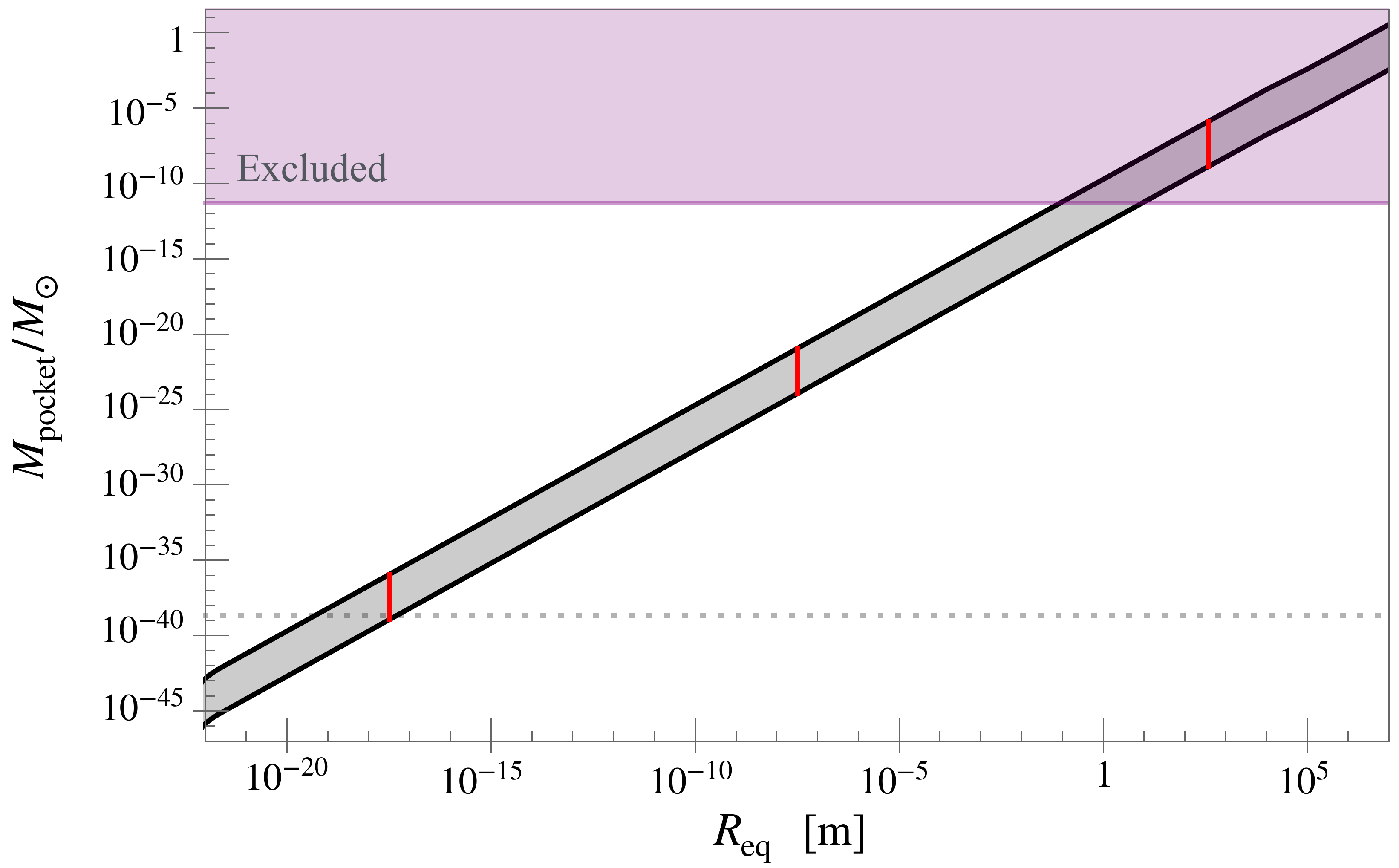}
    \caption{Equilibrium mass and radius of axion relic pockets for $\alpha=1.6$ and $\epsilon=10^{-3}$ (top black line) and $\epsilon=1$ (bottom black line). Vertical red lines, from left to right, respectively indicate $T_t= 10^{10}\, {\rm GeV},~10^{5}\, {\rm GeV} \text{~and~} 1\, {\rm GeV}$.  Maximum temperature used in plot is $T_t = 2\cdot 10^{12}\, {\rm GeV}$. The dotted horizontal line indicates $M_{\rm pocket}=M_{\rm Pl}$. The observationally excluded region is shaded purple, cf.~Sec.~\ref{sec:const} for details.}
    \label{fig:MR}
\end{figure}

Transitions happening very early result in  small pockets that are light but dense and contain hotter gas (in the thermal case). The wide range of predicted radii and masses, shown in Fig.~\ref{fig:MR}, suggest that a broad range of experiment and observations may probe the theory, as we now discuss.

\section{Observational Constraints and Prospects}
\label{sec:const}
\label{sec:obs}
Terrestrial experiments may be sensitive to axion relic pockets produced at very early times so that they remain locally abundant today, cf.~Eqs.~\eqref{eq:Dt0}--\eqref{eq:encounter}. 
Traditional axion haloscope experiments designed to to detect weak radio signals are not sensitive to axion relic pockets. Direct detection experiments designed to search for Weakly Interacting Massive Particles (WIMPs) may be sensitive to axion relic pockets produced close to the maximal temperature, in particular if the axion has additional couplings to matter or electromagnetism.

Astrophysical observations provide a promising route to test the theory. Indeed, existing constraints on the abundance of primordial black holes and MACHOs through their microlensing signal carry over to axion relic pockets. The gravitational lensing of compact astrophysical objects can cause transient flux amplifications of distant, point-like sources (like stars). Constraints on the frequency of such events have been interpreted as limits on the density of compact dark matter objects, assuming the dark matter distribution is sufficiently homogeneous (cf.~\cite{Green:2020jor} for a review).  Current limits from stellar microlensing exclude such compact objects as the dominant source of dark matter over a wide range of masses, $5\cdot 10^{-12} < M/M_\odot < 30$ \cite{Paczynski:1985jf,EROS-2:2006ryy,Macho:2000nvd,Niikura:2019kqi,Croon:2020ouk}.\footnote{We expect this constraint to approximately hold even for the most dilute pockets in this range, as we observe that $R_{\rm eq}$ is smaller than the Einstein radius relevant for microlensing (cf.~e.g.~\cite{Croon:2020ouk,Croon:2020wpr}).}
 For $M\gtrsim 10M_\odot$, this lensing effect would modify the observed distribution of supernova brightness~\cite{Metcalf:2006ms,Zumalacarregui:2017qqd,Garcia-Bellido:2017imq}, and these populations would easily disrupt wide binaries~\cite{Monroy-Rodriguez:2014ula} and cold star clusters~\cite{Brandt:2016aco} as well.
This excludes the low-transition-temperature  region of the axion relic pocket parameter space, roughly corresponding to  $T_t \lesssim 15\,{\rm GeV}$ (cf.~Fig.~\ref{fig:MR}).

Compact dark matter objects could be gravitationally captured by neutron stars~\cite{Capela:2013yf} or white dwarfs~\cite{Graham:2015apa}, triggering instability and leading to constraints on compact objects with mass $M\lesssim 10^{-12}M_\odot$. Such constraints are unlikely to be relevant for axion relic pockets, given that the large magnetic fields produced by neutron stars and white dwarfs would likely lead to axion-photon conversion rather than capture (see below).

Dilaton phase transitions can in principle generate an appreciable  gravitational wave signal from the collision of the bubble walls and the non-spherical initial oscillations of the pockets, cf.~e.g.~\cite{GarciaGarcia:2016xgv} for work on this topic in a similar context to ours. However,  for $T_t \gg 1$ eV, the energy density in the axion-dilaton sector is only a small fraction of the total energy density, cf.~Eq.~\eqref{eq:energyfraction}, and the gravitational wave signal is very weak.

An exciting possibility is to search for smoking-gun signals from the hot axion gas that stabilises the pockets. Axions that couple to electromagnetism through the interaction
\begin{align}
    {\cal L} = - \frac{1}{4} g_{a\gamma}\, a\, F_{\mu \nu} \tilde  F^{\mu \nu} \, ,
\end{align}
can interconvert into photons in background magnetic fields. Here $\tilde  F^{\mu \nu}= \tfrac{1}{2}\epsilon^{\mu \nu \lambda \rho} F_{\lambda \rho}$, and $g_{a\gamma}$ denotes the axion-photon coupling. The axion-photon coupling is expected to depend polynomially on the dilaton and only change by a moderate amount between the true and false vacuum (cf.~e.g.~\cite{Reece:2024wrn} for a review). Observational constraints on  $\gagtv$, reviewed in  \cite{PDG}, are then expected to translate to similar constraints on $\gagfv$. Thus, we don't expect it to be possible to take $\gagfv \gg \gagtv$ in fully explicit models.

Axion relic pockets located in or passing regions with a strong magnetic field are more likely to generate a detectable photon flux.  This suggests that axion relic pockets may generate characteristic astrophysical signals that are morphologically distinguishable from e.g.~those of annihilating or decaying WIMP dark matter.
To estimate the rate at which axions in false-vacuum pockets convert into photons, we recall that the conversion probability for an axion passing through a region of size $L$ with a constant magnetic field component perpendicular to the trajectory, $B_\perp$, is  given by
\begin{align}
    P_{a\gamma} =\frac{\Theta^2}{1+ \Theta^2}\sin^2\left(\Delta \sqrt{1+ \Theta^2}\right) \, , 
    \label{eq:Pag}
\end{align}
where $\Theta=2 g_{a\gamma}B_\perp \omega/m^2_{\rm eff}$ and $\Delta = m_{\rm eff}^2 L/(4 \omega)$ where $\omega$ denotes the energy of the relativistic axion and $m^2_{\rm eff} = m_a^2- \omega_{\rm pl}^2$, where $\omega_{\rm pl}$ 
denotes the plasma frequency (cf.~e.g.~\cite{Raffelt:1987im} for more details).   As discussed in Sec.~\ref{sec:PT}, the mass of the axion in the false vacuum is bounded from below by $\mafv\gtrsim 3H(t_t)$, and, as we have discussed above, the transition temperature is bounded from below by microlensing. The minimal false-vacuum mass that remains of phenomenological interest is then $\text{min}\big( \mafv \Big) \approx 3H(T_t\approx 15{\, \rm GeV}) \approx 10^{-6}$ eV. For environments with free electron densities $n_e \lesssim 10^8~{\rm cm}^{-3}$, the plasma frequency contribution to $m_{\rm eff}^2$ can safely be ignored. Thus $m_{\rm eff}^2 = \masqfv$ in most interesting environments. Finally, as we will return to below, for sufficiently high mode energies, Eq.~\eqref{eq:Pag} is modified by QED birefringence and Cosmic Microwave Background (CMB) dispersion effects \cite{Raffelt:1987im, Dobrynina:2014qba}.

The trapped axions bounce around inside the false-vacuum pocket, and the conversion probability is to a good approximation given by the sum of the conversion probabilities generated between each reflection (interference terms average out). Assuming an isotropic velocity distribution, particles reflected off the pocket wall travel on average a distance $L = 4 R_{\rm eq}/\pi$ before their next reflection. Taking $\omega=T_{\rm eq}$ and using Eq.~\eqref{eq:Tthermal}, the mixing parameters are given by
\begin{align}
    \Theta&= \frac{2\gagfv }{9\tilde m^2} B_\perp  T_{\rm eq} R^2_H(t_t) 
    =
  1.2 \cdot 10^{-12} ~ \frac{\tilde g_{10} \tilde B_{\mu {\rm G}}}{\tilde m^2 \tg^{3/4} \epsilon^{1/4} } \ttt^{-13/4} 
    \, , \nonumber \\
    \Delta &= \frac{9\tilde m^2 R_{\rm eq}}{\pi T_{\rm eq} R^2_H(t_t)} = 
    1.2\cdot 10^{-12}~
    \tilde m^2 (1+\alpha)^{1/3} \epsilon^{1/4} \tg^{1/4} \ttt^{5/4} 
          \, ,
      \label{eq:Theta} 
\end{align}
where $\tilde m= \mafv/3H(t_t)$,
$\tilde g_{10}=\gagfv/(10^{-10}\, \text{GeV}^{-1})$, 
and
$\tilde B_{\mu {\rm G}}= B_\perp/\mu{\rm G}$.
 When both mixing parameters are small, $\Theta \ll1,~\Delta \ll1$, or alternatively, when $\Theta\gg1$ but $\Delta  \Theta \ll1$, the conversion probability simplifies and becomes energy independent: 
\begin{align}
    P_{a\to\gamma} &= \Theta^2 \Delta^2 = \frac{1}{4} g_{a\gamma}^2\big|_{\rm FV} B_\perp^2 L^2 
 \nonumber \\   
    &=  2.0 \cdot 10^{-48}~\tilde g_{10}^2 \,\tilde B_{\mu {\rm G}}^2 \tg^{-1} (1+\alpha)^{2/3}
    \ttt^{-4}  
    \, .
\end{align}
In this approximation, the axion-photon conversion rate is given by 
\begin{align}
    \Gamma_{a\to \gamma} = \frac{P_{a\to \gamma}}{L}=
 6.5\cdot10^{-20}~\text{Gyr}^{-1}~ \tilde g_{10}^2 \tilde B_{\mu {\rm G}}^2\,\tg^{-1/2} (1+ \alpha)^{1/3} \ttt^{-2} \, .
\end{align}
This shows  that axion relic pockets remain stable on cosmological scales, even in the presence of an axion-photon coupling. The number of photons emitted from an axion relic pocket per unit time is given by
\begin{align}
    \left|\frac{d N_{a\, {\rm pocket}}}{dt} \right| = \Gamma_{a\to \gamma} N_{a\, {\rm pocket}} = 
    200~{\rm s}^{-1}~ \tilde g_{10}^2 \tilde B_{\mu {\rm G}}^2 (1+ \alpha)^{5/6} \tg^{-5/4} \epsilon^{-3/4} \ttt^{-23/4}
\, .
\label{eq:flux}
\end{align}
The photon flux is strongly enhanced for low transition temperatures and high magnetic fields. However, Eq.~\eqref{eq:flux}
ceases to be applicable when QED birefringence effects become important, which happens for \cite{Raffelt:1987im}
\begin{align}
\frac{\tilde \omega_{\rm GeV}^2 \tilde B_{\mu {\rm G}}^2}{\tilde m_{\mu {\rm eV}}^2}\gtrsim 1.3 \cdot 10^{13} \, ,
\end{align}
where $\tilde m_{\mu {\rm eV}}= \mafv/\mu\text{eV}$ and $ \omega_{\rm GeV}= \omega/\text{GeV}$. Considering the lowest viable transition temperature and axion mass, $\omega_{\rm GeV} \sim T_{\rm eq}(T_t=15\, {\rm GeV}) \sim 230$ MeV and $\tilde m_{\mu {\rm eV}}=1$, and ignoring factors of $(1+\alpha)$, $\tg$ and $\epsilon$, we find that Eq.~\eqref{eq:flux} holds for $B_\perp \lesssim 15$ G. For this maximal magnetic field, we find $\Theta\gg1$ and $\Delta \Theta \ll1$ so that Eq.~\eqref{eq:flux} is applicable and predicts the flux $\left|d N_{a\, {\rm pocket}}/dt \right|\sim 4\cdot 10^{37}\,  \tilde g_{10}^2\, {\rm s}^{-1}$. Depending on the astrophysical system in question, such a large flux could provide  exciting opportunities to search for axion relic pockets through gamma-ray observations. More strongly magnetised systems, like neutron star magnetospheres that can reach field strengths of up to $10^{15}$ G, fall outside the region of applicability of Eq.~\eqref{eq:flux}, but are also of great observational interest.\footnote{The observational radio signals from \emph{non-relativistic} axion clumps, like axion miniclusters and axion stars, colliding with neutron stars have been studied in \cite{Witte:2022cjj}.} We expect to return to the astroparticle physics of axion relic pockets in future work.

\section{Discussion}
\label{sec:disc}
We have shown that a quantum phase transition of a dilaton field can trap axions into pockets of the false vacuum state, leading to a novel scenario for dark matter, distinct from existing paradigms. 

Throughout this paper, we have assumed that the axion mass difference between the true and the false vacuum is sufficiently large to trap the axions into pockets, even after they are accelerated by the dilaton-bubble walls. 
According to \cite{Lewicki:2023mik}, the trapping is efficient when $\matv \gtrsim {\cal O}(10~T_{\rm eq})$. We can now evaluate what this means in terms of the instanton action and dilaton field displacement. Conservatively, we consider the minimum value $\mafv = 3 H(t_t)$, as a larger false-vacuum axion mass only makes the trapping easier. From Eq.~\eqref{eq:Deltam} we have that
\begin{align}
    \masqtv = \masqfv \, {\rm exp}\Big[ S_{\rm FV}\, \tfrac{|\Delta \phi|}{\phi_{\rm FV}} \Big] 
   \gtrsim 100~ T_{\rm eq}^2 \, ,
\end{align}
which gives the condition
\begin{align}
   S_{\rm FV}\, \tfrac{|\Delta \phi|}{\phi_{\rm FV}} \gtrsim \ln\left(\frac{100}{9} (T_{\rm eq} R_H)^2 \right)= 60.2 - \ln\left(
\tg^{1/2} \epsilon^{1/2} \ttt^{5/2}\right)
   \, .
   \label{eq:hierarchy}
\end{align}
Thus, the required mass hierarchy is readily obtained for moderate fractional field displacements, $|\Delta \phi|/\phi_{\rm FV} \sim {\cal O}(0.2)$, in gauge theories that are weakly coupled in the ultraviolet and generate instanton actions of the order of ${\cal O}(200\text{--}300)$, cf.~Sec.~\ref{sec:dilaton}. Clearly, the sensitivity to the assumed minimal ratio between $\masqtv$ and $T^2_{\rm eq}$ is only logarithmic; larger hierarchies can easily be accommodated. Moreover, in Eq.~\eqref{eq:hierarchy} we have conservatively assumed a thermalised axion gas; axion relic pocket in the non-thermal scenario discussed in Sec.~\ref{sec:pheno} typically relax the required hierarchy.  

Several ingredients of our scenario appear ubiquitously in string compactifications, including hidden sector gauge theories, dilatons (i.e.~moduli) paired with axions, large instanton actions, and geometric hierarchies of couplings motivating relatively focused reheating into the visible sector. Indeed, string compactifications on topologically rich compactification manifolds feature several hundreds of dilaton-axion pairs. Our scenario can be realised if merely one of these dilatons  undergo a quantum phase transition; if multple fields tunnel,  the dark matter density may be built up from several axion relic pocket sectors.\footnote{Since dilaton tunneling events can transform relativistic axions into dark matter through the formation of axion relic pockets, our scenario may be helpful to models that otherwise would overproduce axion dark radiation.} We note that the required dilaton potential can straightforwardly  be engineered in supergravity using the prescription of \cite{Kallosh:2017wnt}. In string theory, asymptotically large dilaton values correspond to decompactification limits, or something similar, in which the theory simplifies and becomes more restrictive. Clearly, the vacuum structure we postulate must be realised at more moderate field values.  

In this work, we have considered a dilaton phase transition of a hidden sector. It is also interesting to explore the potential impact of a dilaton phase transitions involving  Standard Model parameters, e.g.~the QCD coupling (cf.~\cite{Ipek:2018lhm, Croon:2019ugf} for related work). We will return to this question in future work. Moreover, it would be interesting to interface the dilaton phase transition with the thermal Peccei-Quinn phase transition in axion theories with a renormalizable ultraviolet completion within four-dimensional field theory. In particular, when the Peccei-Quinn transition occurs before the dilaton phase transition, this could result in rich dynamics involving both cosmic strings and axion relic pockets.

In this paper, we have developed the axion relic pocket theory  analytically. However, numerical simulations of the phase transition and pocket formation are required to fully determine the predictions of the theory. In particular, the pocket peculiar velocity (cf.~$\epsilon$) and initial gamma-factor of the contracting wall (cf.~$\alpha$) are left as undetermined parameters in our analytical treatment. A full-scale simulation of the phase transition would also shed light on the pocket merger rate: we expect it to be difficult for relativistic pockets to merge, while non-relativistic pockets may already be too dilute for mergers to be relevant. 

Moreover, numerical simulations of individual pockets may determine the damping rate of radial oscillations observed in Sec.~\ref{sec:pockets} (and previously  studied in \cite{Lewicki:2023mik}). Extending such simulations to non-spherical pockets would be very interesting.  

Throughout this paper, we have used the thin-wall approximation to describe the expanding bubbles and contracting pockets. Relaxing this assumption, it becomes possible for axions to partly penetrate the wall before being reflected. As the axions decelerate in the thick wall, they experience rapidly increasing self-couplings, potentially enabling enhanced thermalisation rates and novel phenomena. Moreover, bubble coalescence at $t_{\rm coll}$ involves both large, thin-wall bubbles as well as smaller, thick-wall bubbles, which may impact the coalescence dynamics.

  Depending on the transition time, the size and mass of axion relic pockets can vary by many orders of magnitude, cf.~Fig.~\ref{fig:MR}. We anticipate that laboratory experiments can be designed to probe axion relic pockets forming soon after inflation (say, $T_t\gtrsim 10^{14}$ GeV), while a host of astrophysical signals can be envisioned for phase transitions occurring at lower temperatures. We expect to return to these questions in future work.

We have shown that the highly constrained, exponential interaction between the dilaton and the axion can easily result in large mass hierarchies. We have also shown that the standard misalignment mechanism, guaranteed from inflation, suffices to generate the initial seed population of axions. This makes trapped axions ideal candidates for providing the pressure stabilising the relic pockets. However, many of the predictions of the scenario follow from simple considerations such as pressure and energy balance, and do not rely on the details of axion physics. Other microscopic realisations are also possible; for example, a first-order higgsing of a thermal relic, massless in the false vacuum but heavy in the true vacuum, can result in relic pockets.  For discussions of similar scenarios, see e.g.~\cite{Gross:2021qgx, Kawana:2021tde}.

\section*{Acknowledgements}
We thank Rudin Petrossian-Byrne,  Tim Linden, Filippo Sala, Bo Sundborg, Jorinde van de Vis, Rishav Rohan, Graham White and Sam Witte for stimulating discussions. 
The work of P.C., J.E.~and D.M.~was supported by the Swedish Research Council (VR) under grants 2018-03641 and 2019-02337. The work of O.I. was supported by the European Union's Horizon 2020
research and innovation program under the Marie Skłodowska-Curie grant
agreement No.~101106874.
This article is based upon work from COST Action COSMIC WISPers CA21106, supported by COST (European Cooperation in Science and Technology).

\appendix

\section{Axion pressure on dilaton bubble walls}
\label{app:pressure}
\label{sec:ODE}

Upon reflection, axions exert pressure on  bubble walls. We consider a planar bubble wall surface element of area $\Delta {\cal A}$  extending in the $xy$-plane and moving with velocity ${\bf v}_w = v_w \hat z$ in the cosmological rest frame. 
An axion will gain (or lose) energy when reflected off the wall. We assume the axion is located to the left of the wall in the $z$-direction and denote the initial four-momentum in the cosmological rest frame by
$
p_{\rm in}^\mu = (E,\, {\bf p}) \, .
$
We require $p_{z}/E > v_w$ so that a collision happens.
The axion momentum in the wall's rest frame is given by
$$
\tilde{p}_{\rm in}^\mu = \Big(\gamma_w \left(E- v_w p_{z} \right),\, p_{x},\, p_{y},\, \gamma_w\left( p_{z}- v_w E\right) \Big) \, ,
$$
where $\gamma_w = (1- v_w^2)^{-1/2}$. In the wall frame (and for sufficiently heavy walls), the reflection flips the sign of the $z$-component of the axion momentum. Transforming back to the cosmological rest frame then gives  
$
p_{\rm out}^\mu = (E_{\rm out}, {\bf p}_{\rm out})$, where $p_{x,{\rm out}}=p_{x}$, $p_{y,{\rm out}}=p_{y}$, and
\begin{align}
    E_{\rm out} &= \gamma_w^2\left(E(1+  v_w^2) - 2 v_w p_{z}
\right) \nonumber \\
p_{z,{\rm out}} &= \gamma_w^2\left(-p_{z}(1+  v_w^2) + 2 v_w E
\right) \nonumber \, .
\end{align}
The pressure exerted on the wall over the time period $\Delta t$ equals the total momentum change per unit area and unit time,
\begin{align}
    {\cal P} = \sum_{\rm collisions} \frac{- (p_{z, {\rm out}}- p_{z})}{\Delta {\cal A}\cdot \Delta t} = \sum_{\rm collisions} \frac{2 \gamma_w^2(p_{z}- v_w E)}{\Delta {\cal A}\cdot \Delta t}
\end{align}

It is instructive (but, as we will show, for our purposes ultimately insufficient) to seek to describe the evolution of the trapped axions with locally spatially homogeneous and isotropic statistical distribution functions, i.e.~$f(p)$ independent of ${\bf x}$. For example, the following derivation applies to systems in local thermal equilibrium.  The pressure exerted by the axions on the wall can then be expressed as (cf.~e.g.~\cite{Kawana:2022lba, Lewicki:2022nba}),
\begin{align}
    {\cal P} &= 
    \int \frac{d^3{\bf p}}{(2\pi)^3}
   f(p^\mu) \left(\frac{p_{z}}{E} -v_w \right)\cdot 2 \gamma_w^2(p_{z}- v_w E) \cdot \Theta\big(\frac{p_z}{E} -v_w\big)
    \\ \nonumber
    &=
\frac{4 \pi \gamma_w^2}{(2\pi)^3} \int_0^\infty dp\,  f(p) \frac{p^2}{E}
\int_{x_{\rm min}}^1 dx \Big(px-v_wE\Big)^2
    =
    \frac{4 \pi \gamma_w^2}{(2\pi)^3} \int_0^\infty dp\,  f(p) \frac{p(p-E v_w)^3}{3 E} \, ,
\end{align}
where $x_{\rm min}= {\rm max}(0,\, E v_w/p)$.

We first consider the non-relativistic limit motivated by a bubble moving through an axion background produced e.g.~through the misalignment mechanism. For $E |v_w| \gg p$, the pressure simplifies to 
\begin{align}
    {\cal P} =
   - \frac{m_a^2 v_w^3 \gamma_w^2}{6\pi^2} \int_0^\infty dp\,  f(p) p =
    \frac{\delta^2  |v_w|^3 \gamma_w^2}{12\pi^2} m_a^4\, ,
\end{align}
where, since the axions are located to the left of the wall and are approximately stationary,  $v_w\lesssim0$ is required for collisions to happen, and, in the last step, we have assumed a uniform distribution in $p$ up to $\delta \cdot m_a$. While this distribution is motivated purely by mathematical simplicity, we note that the pressure from non-relativistic axions in theories where $m_a^4/\Delta V \ll1$ is highly suppressed for all but extremely large wall gamma-factors.

Considering now  the ultra-relativistic limit, $E = p$, the pressure experienced by the wall is given by
\begin{align}
    {\cal P} = \gamma_w^2 (1-v_w)^3 \frac{4\pi}{3(2\pi)^3 } \int_0^\infty dp\, f(p) p^3
    = \gamma_w^2 (1-v_w)^3\, P_{\rm gas} = \frac{(1-v_w)^2}{1+v_w}\, P_{\rm gas}
    \, ,
    \label{eq:Prel}
\end{align}
where $P_{\rm gas}$ denotes the internal  pressure in the gas pressure,
\begin{align}
    P_{\rm gas} = \frac{1}{(2\pi)^3} \int d^3{\bf p}\, f(p) \frac{p^2}{3E} \, .
\end{align}
For any relativistic (and homogeneous) distribution function, $f(p)$, the internal pressure is  related to the energy density as $P_{\rm gas}=\rho_{\rm gas}/3$. Moreover, the total energy of the gas changes due to the wall pressure as
\begin{align}
    \frac{dE_{\rm gas}}{dt} = 4\pi R^2(t) v_w {\cal P} = \frac{v_w}{R} \frac{(1 -v_w)^2}{1+v_w} E_{\rm gas} \, ,
\end{align}
where $R$ denotes the radius of the wall. Re-expressing this equation  for $P_{\rm gas}(t)$, we find a closed set of coupled ordinary differential equations governing the co-evolution of the wall and the gas pressure:
\begin{align}
    \begin{cases}
        \ddot{R} + 2\frac{1-\dot{R}^2}{R} &= \frac{(1-\dot R^2)^{3/2}}{\sigma}\left( - \Delta V + P_{\rm gas} \frac{(1 - \dot R)^2}{1 +\dot R} \right)\\
        \frac{d}{dt}\ln P_{\rm gas} &= - \frac{\dot R}{R} \left(3 + \frac{(1  - \dot R)^2}{1 + \dot R} \right)
    \end{cases} \, .
    \label{eq:ODEs}
\end{align}
This is a new result of this work which we expect to be very useful for systems with moderate wall velocities. However, in the theory of axion relic pocket formation, the wall gamma-factors can reach very large values. Our second equation of Eq.~\eqref{eq:ODEs} then predicts an exponential increase in the gas pressure at a time-scale of $\sim R/\gamma^2_w$. This is unphysical: causality limits the time scale for a homogeneous pressure to increase to $\gtrsim R$. The rapid pressure increase reflects the break-down of the assumption of homogeneity. This is consistent with the observed formation of shock fronts in our example-simulation of Fig.~\ref{fig:pocket_sim}. Thus, to analyse axion relic pockets, one must use methods that do not rely on homogeneity.  

\bibliographystyle{bibi}
\bibliography{bib.bib}

\end{document}